\documentclass[preprintnumbers,showpacs,amsmath,amssymb,floatfix,prd,onecolumn,superscriptaddress,nofootinbib]{revtex4}
\usepackage{graphicx}
\usepackage{epsfig}
\usepackage{bm}
\usepackage{amsfonts}
\usepackage{epstopdf}
\usepackage{bm}
\usepackage{setspace}
\usepackage{neuralnetwork}
\begin{document}

\title{Multi-pole Dark Energy }

\author{Chao-Jun Feng}
\email{fengcj@shnu.edu.cn} 
\affiliation{Division of Mathematical and Theoretical Physics, Shanghai Normal University,}
\affiliation{Center for Astrophysics, Shanghai Normal University,\\ 100 Guilin Road, Shanghai 200234, China}

\author{Xiang-Hua Zhai}
\email{zhaixh@shnu.edu.cn} 
\affiliation{Division of Mathematical and Theoretical Physics, Shanghai Normal University,}
\affiliation{Center for Astrophysics, Shanghai Normal University,\\ 100 Guilin Road, Shanghai 200234, China}

\author{Xin-Zhou Li}
\email{kychz@shnu.edu.cn} 
\affiliation{Center for Astrophysics, Shanghai Normal University,\\ 100 Guilin Road, Shanghai 200234, China}

\begin{abstract}
While a scalar field with a pole in its kinetic term is often used to study the cosmological inflation, it can also play the role of dark energy, which is called the pole dark energy model. We propose a generalized model that the scalar field may have two or even multiple poles in the kinetic term and we call it the multi-pole dark energy. We find the poles can place some restrictions on the values of the original scalar field with non-canonical kinetic term. After transforming to the canonical form, we get a flat potential for the transformed new scalar field even if the original field has a steep  one. It The late-time evolution of the universe is obtained explicitly for the two pole model, while dynamical analysis is performed for the multiple pole model. We find that it does have a stable attractor solution, which corresponds to the universe dominated by the potential  of the scalar field.
\end{abstract}

\maketitle

\section{Introduction}

Dark energy, the source of the late-time accelerating expansion, has been studied a lot since the observations of Supernovae Type Ia. For example, in Ref.\cite{Cheng:2018nhz}, the authors constructed the type Ia supernova spectrum by training  an artificial neural network, in Ref.\cite{Feng:2009jr}, bulk viscosity of dark energy is taken into account to alleviate the age problem of the universe, and in Ref.\cite{Feng:2009hr}, dark energy is investigated in the braneworld scenario to avoid the big rip ending of the universe, for reviews on dark energy see Refs.\cite{Li:2012dt} and \cite{Bahamonde:2017ize}. It has been shown that more than two thirds of the energy density in the universe is completely unknown, after which the dark energy is named. What we have known is that the equation of state of dark energy is nearly $w\sim-1$ at present \cite{Aghanim:2018eyx}. The gravity force raised by dark energy is a kind of repulsive force; however, in the earth no one has  observed such anti-gravity force in the lab. 

The vacuum energy from  quantum field theory or the cosmological constant from general relativity can be considered as such a kind of dark energy, for its equation of state $w=-1$. However, observations have not confirmed $w=-1$ while actually it deviates from $-1$ slightly\cite{Aghanim:2018eyx}, which means a dynamical dark energy with varying $w$ may be more consistent with observations. Quintessence is a kind of dynamical dark energy model, in which a scalar field minimally coupled to gravity  drives the universe to accelerate. Recently, a scalar field  that  was once used to make the universe inflate in the early time \cite{Kallosh:2013hoa}\cite{Kallosh:2013yoa}\cite{Galante:2014ifa}\cite{Carrasco:2015uma} with a pole in its kinetic term  is proposed as a new kind of dark energy model\cite{Linder:2019caj}, which is called the pole dark energy model. In this model, the original field that is non-minimally coupled to the gravity does not need a very unnatural flat potential. The transformation that transforms the original field with non-canonical kinetic term into a new one with canonical kinetic term could make the potential of the new field much more flat. Then the universe will be accelerated by this flat potential energy.  

We generalize the pole dark energy model and propose a multi-pole one, in which the kinetic term may have multiple poles.  Poles can come from the super-gravity theory due to the nonminimal coupling to the gravitational field or the geometric properties of the K\"ahler manifold\cite{Broy:2015qna}\cite{Terada:2016nqg}. Besides, the k-essence model \cite{ArmendarizPicon:2000dh} is a dark energy model with non-canonical kinetic terms. Here, we treat it phenomenologically as what has been done in Ref.\cite{Linder:2019caj}.  We find the poles can place some restrictions on the values of the original scalar field, which means  the original scalar field does not need to change a lot when its corresponding transformed field with canonical kinetic term  have much more changes. The later time evolution of the universe is obtained explicitly for the two pole model, while dynamical analysis is performed for the multiple pole model. We find that it does have a stable solution, which corresponds to the universe dominated by the potential energy of the scalar field.

In Sec.\ref{sec:mul}, we introduce the multi-pole dark energy model. The relation between the original scalar field that  has two poles in its kinetic term  and the transformed canonical one will be shown, and the properties of the transformed potential will  also be presented. The cosmological evolution driven by the two pole model will be given in Sec.\ref{sec:cos}. For a general multi-pole dark energy, we will perform the dynamical analysis in Sec.\ref{sec:dyn}. In Sec.\ref{sec:dis} discussions and conclusions will be presented.

\section{The multi-pole dark energy }\label{sec:mul}

In general, the Lagrangian for a scalar field with poles in the kinetic term could be written as 
\begin{eqnarray}
\mathcal{L} = -\frac{1}{2}\frac{k^2}{f^2(\sigma)}(\partial\sigma)^2 - V(\sigma)\,,
\end{eqnarray}
where $V(\sigma)$ is the  potential and $f(\sigma)$ is some function of the scalar field. Function $f$ may have multiple zero points by construction. The parameter $k$ could be positive or negative. When $k<0$, it is equivalent to changing the overall sign of the $f$ function while keeping $k$ positive. Without losing of generality, $k$ will be taken as $\pm1$ in the numerical calculations.    Poles can come from the super-gravity theory due to the nonminimal coupling to the gravitational field or the geometric properties of the K\"ahler manifold. In the pole dark energy model Ref.\cite{Linder:2019caj}, function $f$ is taken as a power law: $f(\sigma)=\sigma^{p/2}$, and the pole resides at only one point $\sigma=0$ with residue $k^2$ and order $p$.

\subsection{Two poles}

After performing the  transformation $d\phi = kd\sigma/f(\sigma)$, the non-canonical kinetic term of $\sigma$ is transformed to the canonical form for the scalar field $\phi$ :
\begin{equation}\label{eqi:can}
\mathcal{L} = -\frac{1}{2}(\partial \phi)^2 - V(\sigma(\phi))\,.
\end{equation}
If the function $f$ can be phenomenologically  taken as the following form : 
\begin{eqnarray}\label{equ:mp}
f(\sigma) = \sigma(1 -\beta\sigma^q) \,,
\end{eqnarray} 
which residues at $\sigma=0$ and $\sigma=\beta^{-1/q}$ with parameters $q>1, \beta>0$ in the unit of $8\pi G = 1$, we can get an explicit relation between $\phi$ and $\sigma$:
\begin{eqnarray}
\phi &=& 
\frac{k}{q}\ln\bigg(\frac{\sigma^q}{1-\beta\sigma^q} \bigg)\,, \\
\sigma &=& \left(\frac{1}{e^{-q\phi/k}+\beta } \right)^{1/q} \,.\label{equ:sigp}
\end{eqnarray}
 When $k>0, \beta=0$,  we have $f=\sigma$ from Eq.(\ref{equ:mp}). That is just the pole dark energy model in Ref.\cite{Linder:2019caj} with $p=2$ there. This model is often used for inflation. When $k<0$,  $f\sim \beta \sigma^{q+1}$ for large $\sigma$, which is coincident with the pole dark energy model when $p=2{q+1}$. Function $f$ can be also written in terms of $\phi$: 
 \begin{equation}
 f = e^{\phi/k}\bigg( 1+\beta e^{q\phi/k}\bigg)^{-1-\frac{1}{q}} \,.
 \end{equation}
We will take the branch $0<\sigma<\beta^{-1/q}$, which corresponds to $\phi\in(-\infty, \infty)$. By contrast, $\sigma$ is taken in the branch of $0<\sigma $ in the pole dark energy model. It shows the second pole makes a constraint on the $\sigma$ field. When the parameter $q$ is chosen to be $q<-1$, it takes the branch $\sigma>\beta^{-1/q}$ correspondingly, see Eq.(\ref{equ:sigp}). Therefore, when  two poles are very close to each other, such as a very large $\beta$ in the two pole model,  one can  take another branch by setting suitable values of the parameters, such as $q<-1$ here and the result will not be changed.

In the case of power law potential, we have
\begin{eqnarray}
V \sim \sigma^n \,, \rightarrow V \sim (\beta + e^{-q\phi/k})^{-n/q} \,.
\end{eqnarray}
For $k>0$, when $\phi$ goes to infinite, the potential becomes
\begin{eqnarray}
V|_{\phi\rightarrow\infty} \sim \beta^{-\frac{n}{q}}\left(1-\frac{n}{q\beta}e^{-q\phi/k}\right)\,,
\end{eqnarray}
which is basically an uplifted exponential potential. Otherwise, when $\phi$ goes to minus infinity, the potential becomes $V|_{\phi\rightarrow-\infty}\sim e^{n\phi/k}$. For $k<0$, the limits are exchanged. Note that after transforming to the canonical form, we get a flat potential for the transformed new scalar field $\phi$ even if the original field $\sigma$ has a steep  one. The first derivative of the potential  with respect to $\phi$  is given by:
\begin{eqnarray}
\frac{V_\phi}{V} \equiv \frac{dV/d\phi}{V}  =\frac{n}{k} \frac{ e^{-q\phi/k}}{\beta + e^{-q\phi/k}} \,.
\end{eqnarray}
When $\beta = 0$, $V_\phi/V$ is a constant. However, in the case of  $\beta \neq 0$, $V_\phi/V \sim e^{-q\phi/k}$. For $k>0$, when $\phi$ is going large, $V_\phi/V\ll 1$, which indicates the potential has a flat plateau at that moment.

In the case of a dilaton potential, we have
\begin{eqnarray}
V\sim e^{-\alpha\sigma}\,,\rightarrow V\sim e^{-\alpha (\beta + e^{-q\phi/k})^{-1/q} }\,,
\end{eqnarray}
which gives a super-exponential behavior as that in Ref.\cite{Linder:2019caj}. And $V_\phi/V$ is 
\begin{eqnarray}
\frac{V_\phi}{V} = \frac{\alpha}{k} \frac{ e^{-q\phi/k}}{\beta + e^{-q\phi/k}} \frac{1}{(\beta + e^{-q\phi/k})^{1/q}}\,.
\end{eqnarray}
When $\beta=0$, $V_\phi/V \sim e^{\phi/k}$, and while $\beta\neq 0 $ and $k>0$, $V_\phi/V \sim e^{-q\phi/k}$, which also gives a flat plateau-like potential. 

In fact, for a general potential $V(\sigma)$, we have
\begin{eqnarray}
\frac{V_\phi}{V} = \frac{V_\sigma}{V}\frac{f}{k} = \frac{V_\sigma}{V} \frac{e^{-q\phi/k}}{(\beta + e^{-q\phi/k})^{1/q+1}}\,. \label{equ:vvpoles}
\end{eqnarray}
When $\phi\rightarrow \infty$, $\sigma$ approaches its second residue $\sigma\rightarrow \beta^{-1/q}$ from Eq.(\ref{equ:sigp}). Then $V_\sigma/V$ becomes a constant, and $V_\phi/V\sim e^{-q\phi/k}$. So the second pole in the kinetic term of $\sigma$ can really help us to get a flat plateau-like potential without fine-tuning any parameters.

\subsection{Multiple poles}
When the kinetic term of $\sigma$ has multiple poles, we can not get an analytical formula for $V(\phi)$; therefore, we will perform the dynamical analysis for this general case in Sec.\ref{sec:dyn}. Note that we always take a branch of $\sigma$ that does not cross the residue points. For example, $\sigma\in(0,\beta^{-1/q})$ in the last section. It means the zeros of $f(\sigma)$ will place some restrictions on the values of $\sigma$. The change of $\sigma$ field is then not too much during the evolution, even there is a big change of the $\phi$ field. With help of poles in the kinetic term of $\sigma$ or zero points of the function $f(\sigma)$, we can get a flat plateau-like potential easily, since $V_\phi/V\rightarrow 0$ when $f$ approaches any one of its zero points, see Eq.(\ref{equ:vvpoles}). 

\section{Cosmological evolution}\label{sec:cos}

The late-time evolution of a flat universe is determined by the Friedmann equation :
\begin{equation}了\label{equ:frid}
H^2 = \frac{1}{3M_p^2}\left(\rho_m + \frac{1}{2}\dot\phi^2 + V(\phi)\right) \,,
\end{equation} 
which includes the dark matter and dark energy components. Here $M_p^2=1/8\pi G$ is the reduced Planck mass, and also $M_p=1$  in the unit of $8\pi G =1$. The dot over $\phi$ denotes the derivatives with respect to time, and $\rho_m$ is the energy density of dark matter. The  equation of motion for the $\phi$ field is  given by 
\begin{equation}\label{equ:eom}
\ddot \phi + 3H\dot \phi +\frac{dV}{d\phi} = 0\,.
\end{equation} 
We also have the  dynamic equation:
\begin{equation}\label{equ:deom}
\dot H = -\frac{1}{2} (\rho_m + \dot\phi^2)\,.
\end{equation} 
Let $x=\ln a$ and introduce the following field and potential: 
\begin{equation}\label{equ:rede}
\psi = \frac{\phi}{M_p}\,, \quad  U = \frac{V}{3H_0^2M_p^2}\,,
\end{equation}
where we have recovered the unit to see that both $\psi$ and $U$ are dimensionless and $H_0$ is the present value of the Hubble parameter. Then, the Friedmann equation then becomes
\begin{equation}
E^2 \left( 1-\frac{1}{6}\psi'^2\right)= \Omega_{m0} e^{-3x} + U \,,
\end{equation}
where the prime  denotes the derivatives with respect to  $x$ and   $\Omega_{m0}\equiv \frac{\rho_{m0}}{3H_0^2}$, $E\equiv H/H_0$.  Eq.(\ref{equ:deom}) becomes 
\begin{eqnarray}\label{equ:deom2}
EE' = -\frac{3}{2}\Omega_{m0} e^{-3x} + \frac{1}{2} E^2\psi'^2\,.
\end{eqnarray}
After a straightforward calculation, the equation of motion for $\psi$ could be written as
\begin{equation}\label{equ:eom2}
\bigg(\Omega_{m0} e^{-3x} + U\bigg)\left( \psi'' + \frac{1}{2} \psi'^3 +3\psi'\right)  +3\left( 1-\frac{1}{6}\psi'^2\right)\left(\frac{dU}{d\psi}-\frac{1}{2}\Omega_{m0} e^{-3x}\psi'\right) = 0\,.
\end{equation}
The equation of state  is given by 
\begin{eqnarray}\label{equ:eos}
w &=& \frac{\dot\phi^2/2-V}{\dot\phi^2/2+V} =  -1+2\left[1+\frac{U\left( 6-\psi'^2\right)}{(\Omega_{m0} e^{-3x} + U)\psi'^2}\right]^{-1} \,.
\end{eqnarray} 
It is clear that when the $\psi$ field's kinetic energy  is much smaller than its potential, $w\sim -1$. The evolution of the filed $\psi$ can be obtained by  numerically solving Eq.(\ref{equ:eom2}). 

\begin{figure}[h]
	\begin{center}
		\includegraphics[width=0.45\textwidth,angle=0]{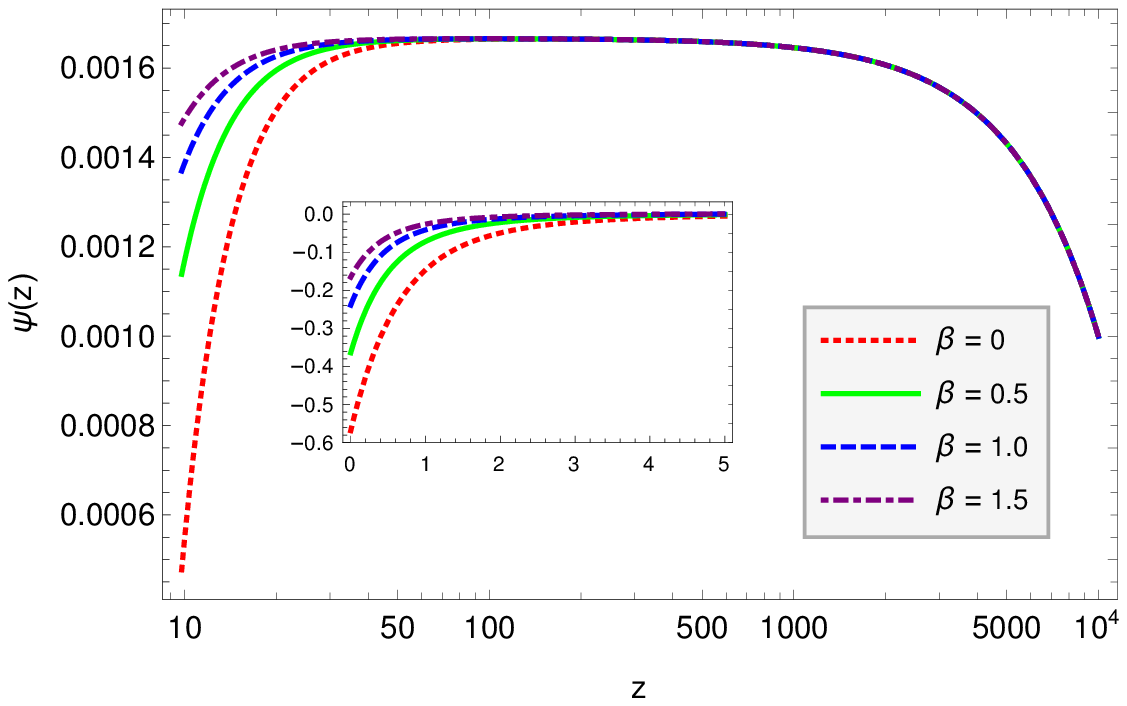}
		\includegraphics[width=0.45\textwidth,angle=0]{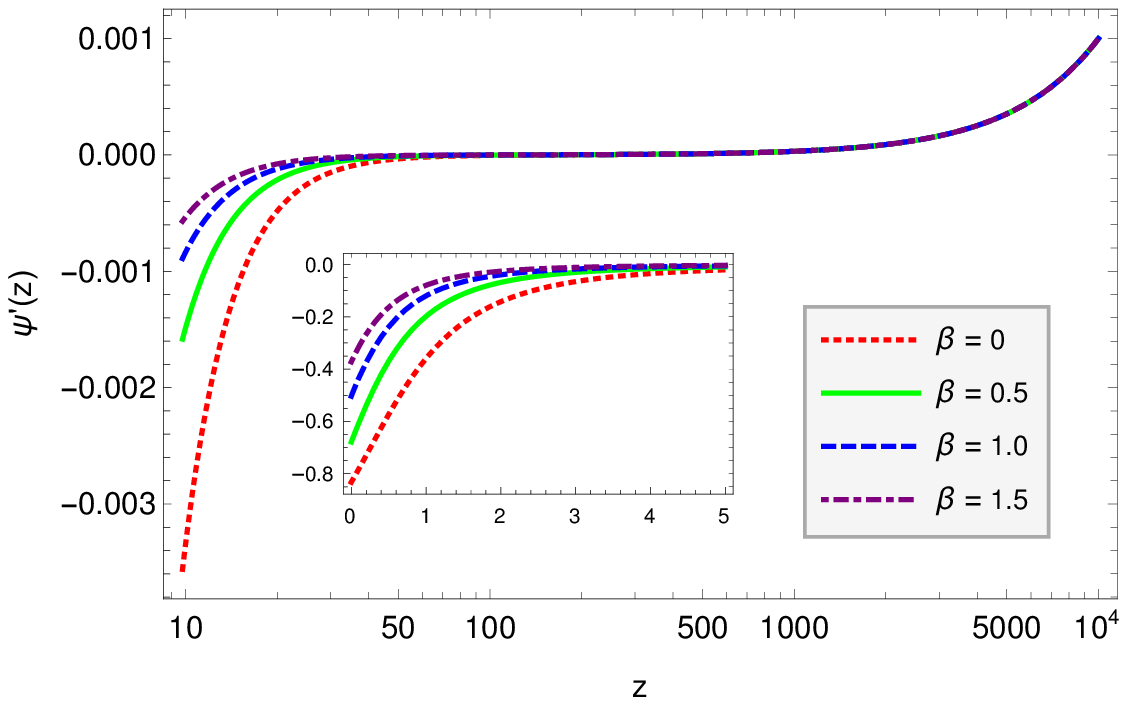}
		\caption{\label{fig:phiz}The evolution of $\psi$ and  $\psi'$ with pow law potential as the function of redshift $z$ with the same initial conditions and with $\Omega_{m0}=0.3,q=2,\alpha=1$ and different $\beta$ values }
	\end{center}
\end{figure}

For $V=m^2\sigma^2/2$ with 
\begin{equation}\label{equ:mod1}
U = U_0(\beta + e^{-q\psi/k})^{-2/q}\,,\quad U_0 = \frac{m^2}{6H_0^2}\,,
\end{equation}
and  $q=2$ we solved Eq.(\ref{equ:eom2}) and plotted the evolution of $\psi$ as the redshift $z=1/a-1$ in Fig.\ref{fig:phiz}. One can always make $k^2=1$ by redefining $\sigma$. Therefore, without losing generality, we set $k=1$ in the numerical calculation. Note that $k=-1$ is equivalent to changing the overall sign of the function $f$  while keeping $k=1$, see Eq.(\ref{equ:mp}). In other words, when $k=-1$, we can redefine $\psi\rightarrow -\psi$, then $U$ will not be changed and $dU/d\psi \rightarrow -dU/d\psi$, and then Eq.(\ref{equ:eom}) will be not changed but an overall minus sign. 

\begin{figure}[h]
	\begin{center}
		\includegraphics[width=0.44\textwidth,angle=0]{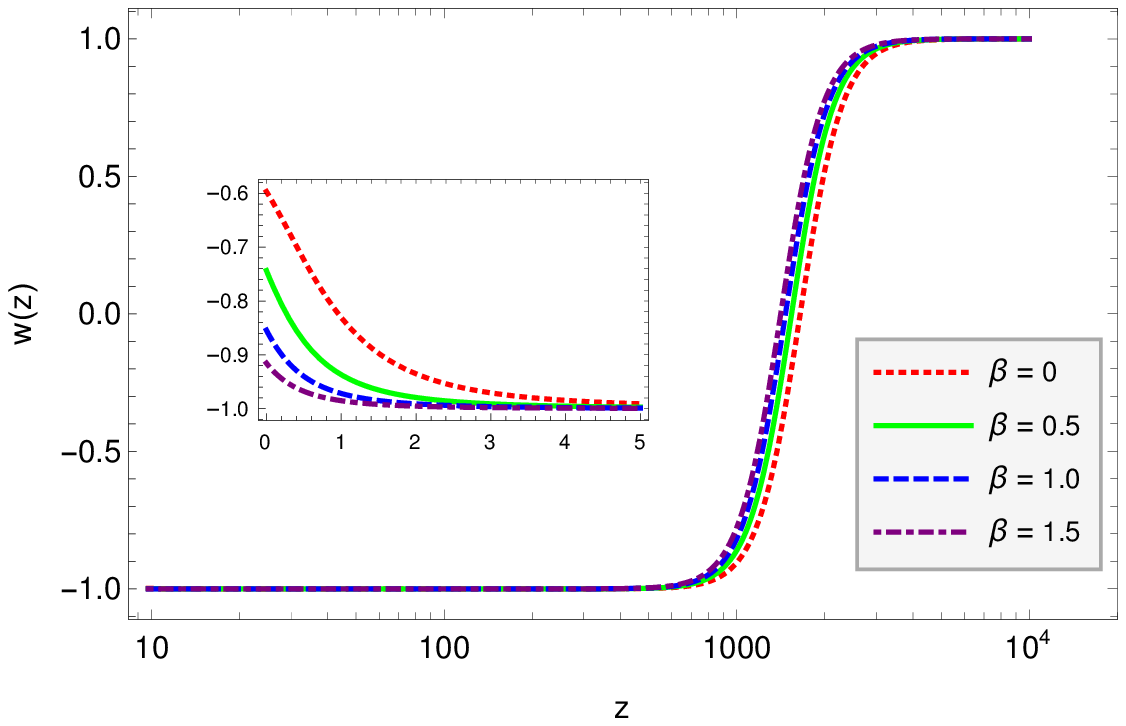}\,
		\includegraphics[width=0.44\textwidth,angle=0]{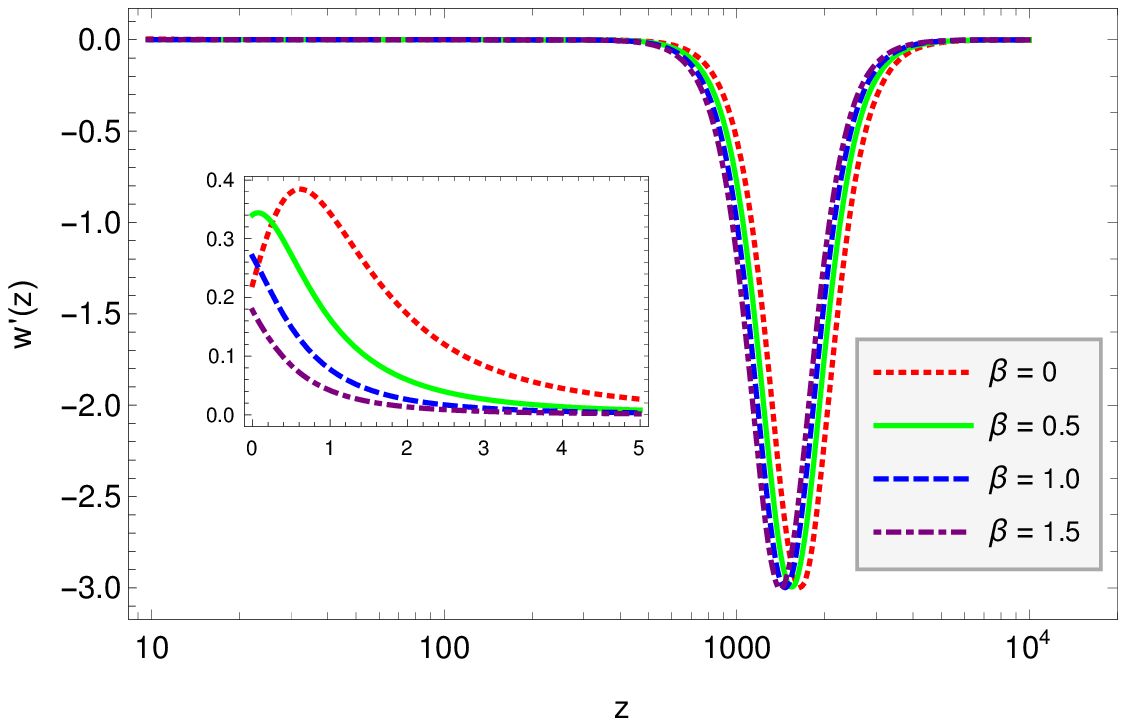}
		\caption{\label{fig:w}For power law potential, the evolution of the equation of state  $w$ and its running $w'$  as the function of redshift $z$ with $\Omega_{m0}=0.3,q=2,n=2$ and different $\beta$ values.   }
	\end{center}
\end{figure}

From Fig.\ref{fig:phiz},  $\psi$ increases in the early time and decreases at present $z\rightarrow 0$. It shows that large value of $\beta$ could slow down the decreasing process of $\psi$, and depress the increasing of the kinetic momentum energy $\sim \psi'^2$. It means that with the help of $\beta$, the potential energy will be the main part of the energy of $\psi$; therefore, its equation of state $w\rightarrow-1$, see Fig.\ref{fig:w}.
\begin{figure}[h]
	\begin{center}
		\includegraphics[width=0.45\textwidth,angle=0]{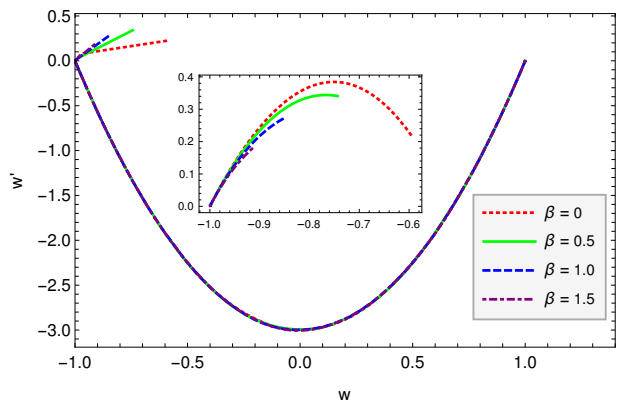}
		\caption{\label{fig:wwz} For power law potential, the dynamics of $w-w'$ phase space with $\Omega_{m0}=0.3,q=2,n=2$ and different $\beta$ values.  }
	\end{center}
\end{figure}

The evolution of the equation of state $w$ and its running $w'$ are plotted in Fig.\ref{fig:w}. We also plot their phase space in Fig.\ref{fig:wwz}.  It is clear that large values of $\beta$ could indeed make the model much more suitable to describe the present accelerating universe. And the running of $w$ almost vanishes ($w'\sim 0$) at present when $\beta$ is large.

\begin{figure}[h]
	\begin{center}
		\includegraphics[width=0.45\textwidth,angle=0]{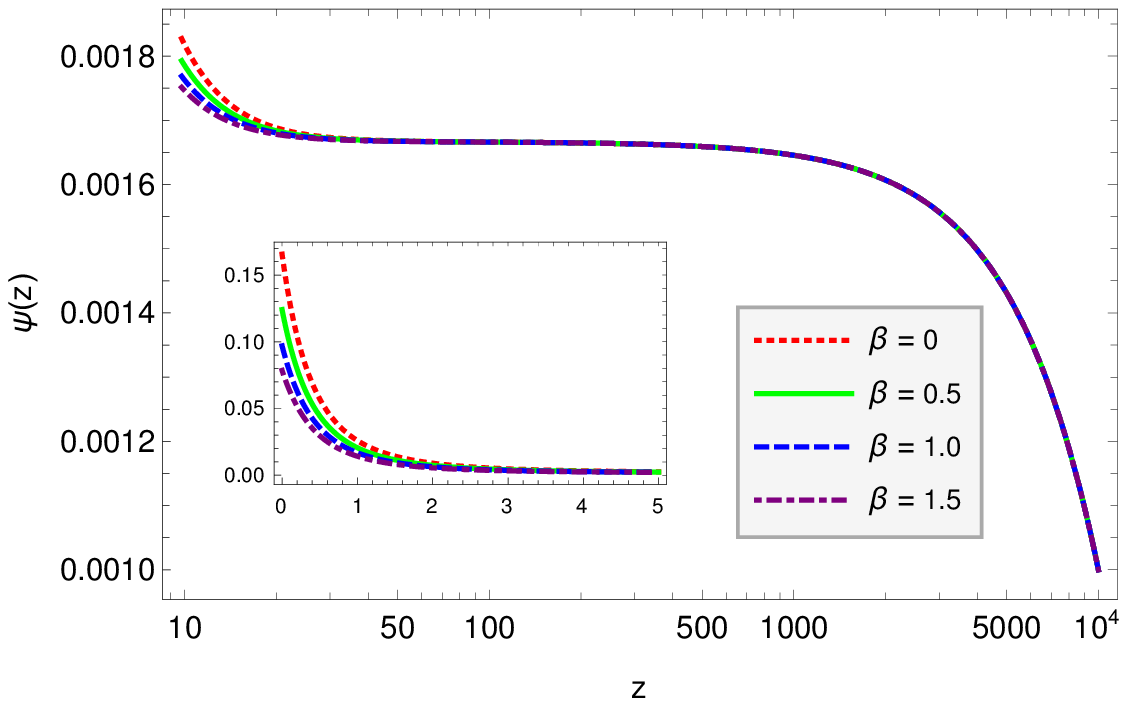}
		\includegraphics[width=0.45\textwidth,angle=0]{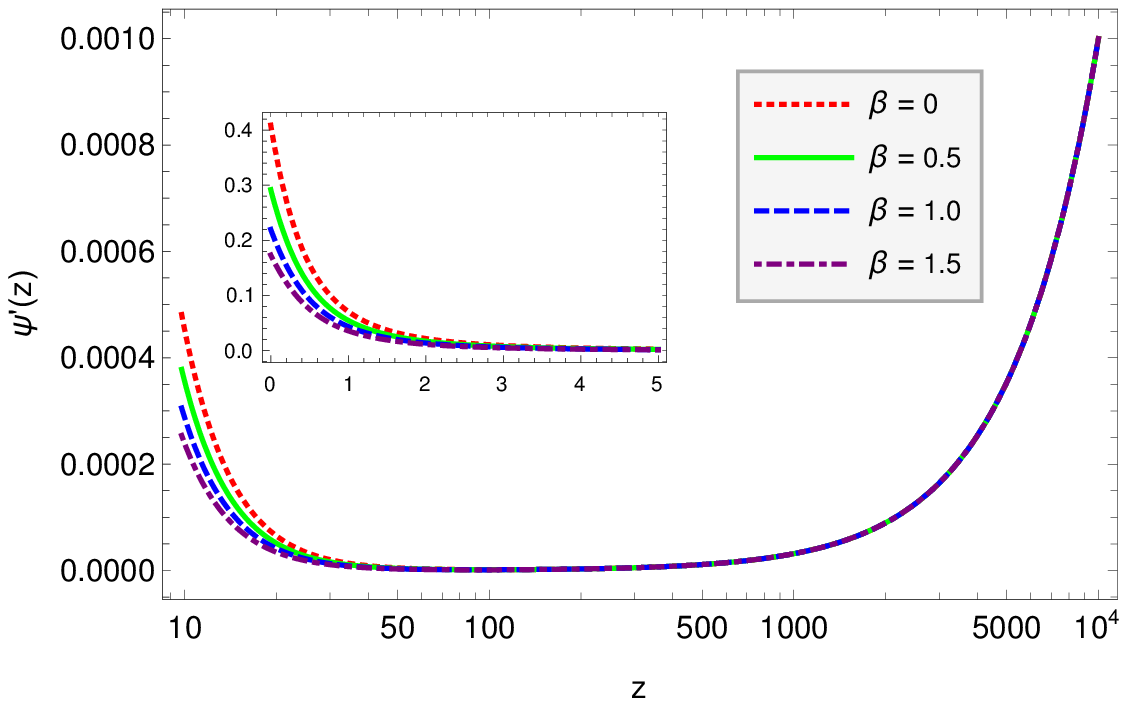}
		\caption{\label{fig:phiz3} The evolution of $\psi$ and  $\psi'$ with dilaton potential as the function of redshift $z$ with the same initial conditions and with $\Omega_{m0}=0.3,q=2,\alpha=1$ and different $\beta$ values.    }
	\end{center}
\end{figure}

\begin{figure}[h]
	\begin{center}
		\includegraphics[width=0.45\textwidth,angle=0]{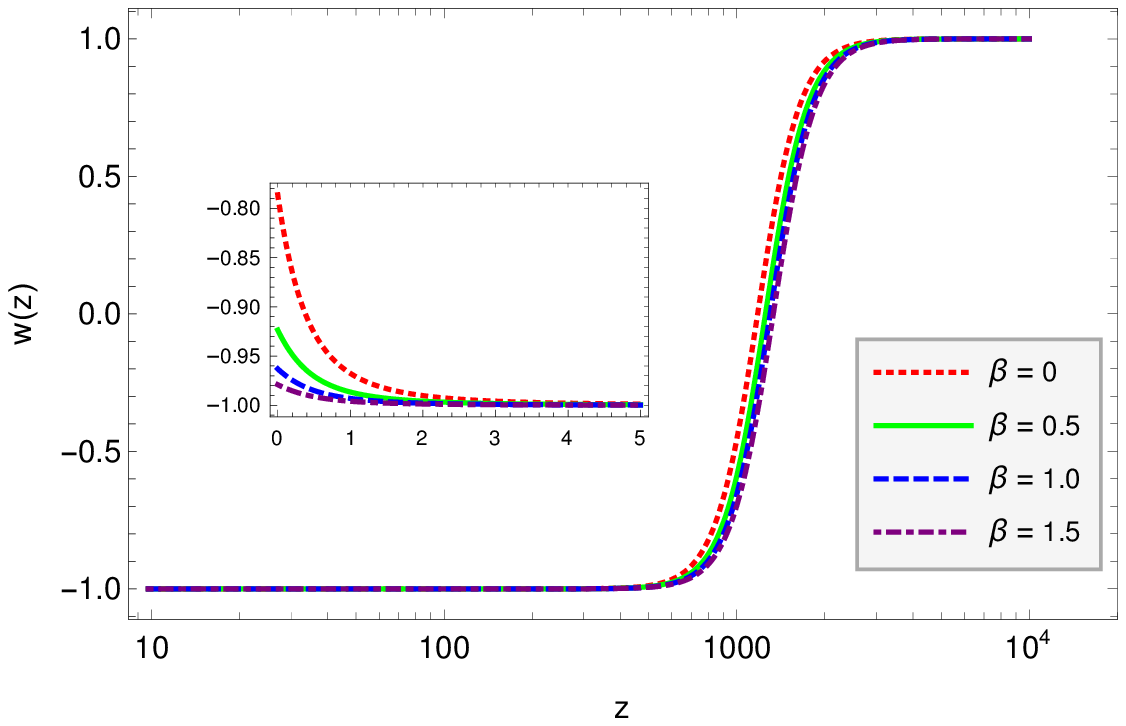}
		\includegraphics[width=0.45\textwidth,angle=0]{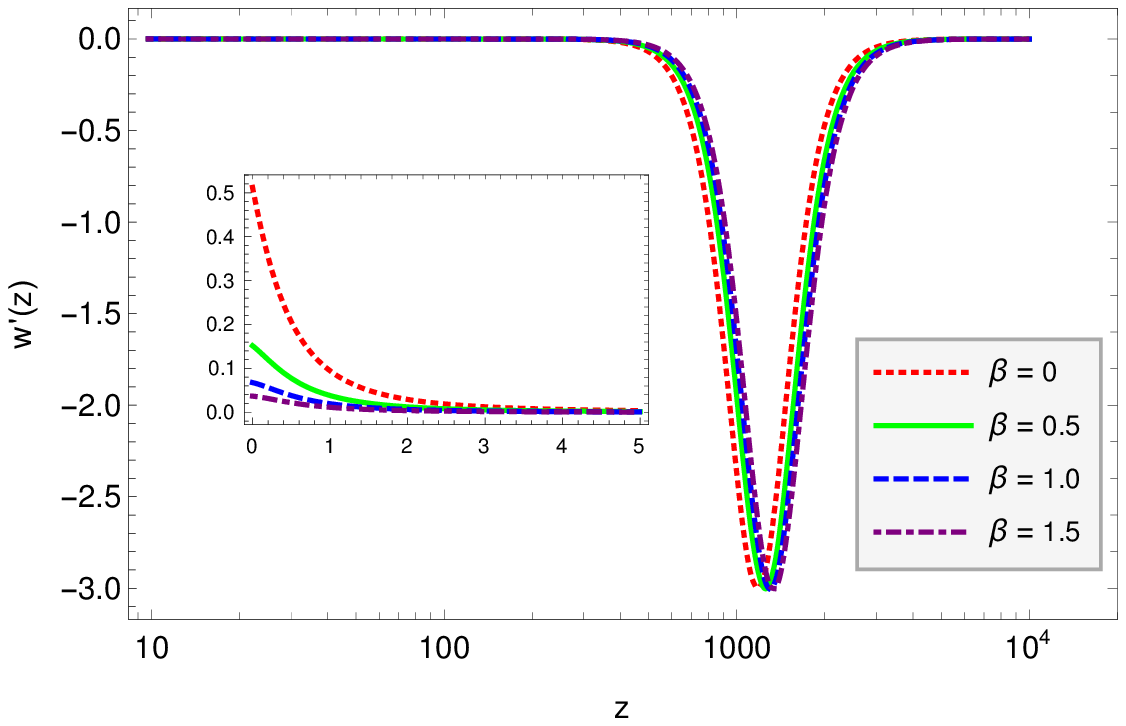}
		\caption{\label{fig:w3}For dilaton potential, the evolution of the equation of state  $w$ and its running $w'$  as the function of redshift $z$ with $\Omega_{m0}=0.3,q=2,\alpha=1$ and different $\beta$ values.   }
	\end{center}
\end{figure}

\begin{figure}[h]
	\begin{center}
		\includegraphics[width=0.45\textwidth,angle=0]{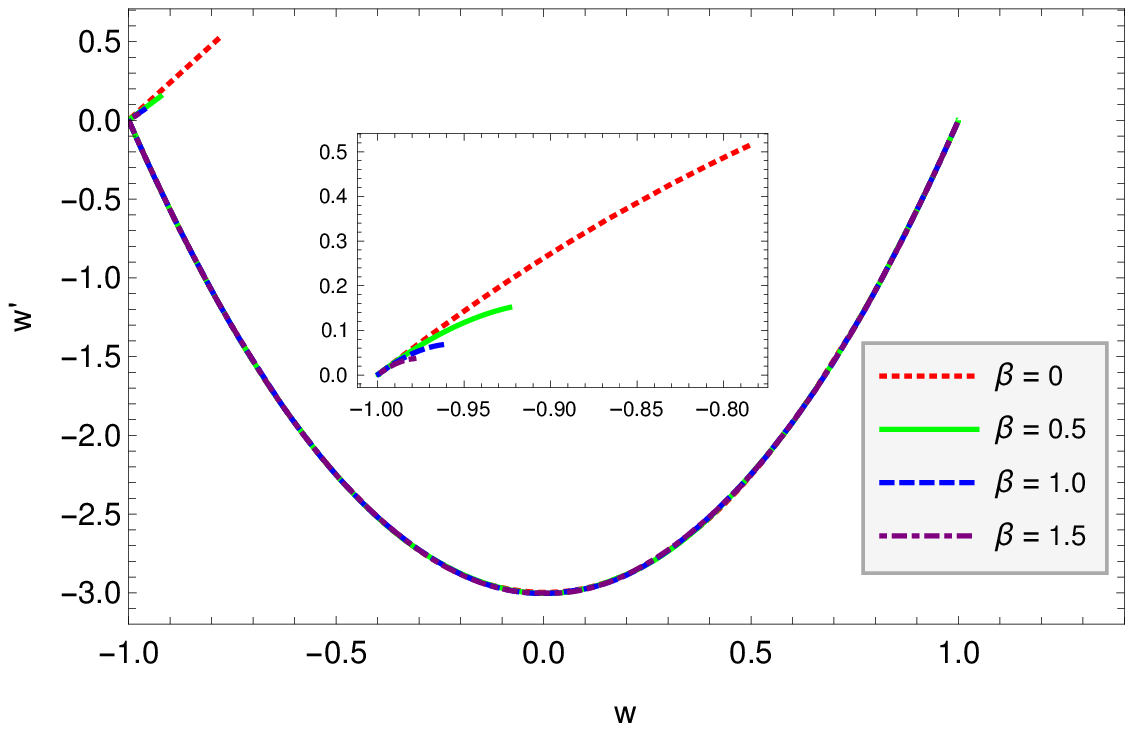}
		\caption{\label{fig:wwz3} For dilaton potential, the dynamics of $w-w'$ phase space with $\Omega_{m0}=0.3,q=2,\alpha=1$ and different $\beta$ values.  }
	\end{center}
\end{figure}

Now we take the potential as $V= V_0e^{-\alpha \sigma}$, or
\begin{equation}
U = U_0e^{-\alpha (\beta + e^{-q\psi/k})^{-1/q} } \,,\quad  U_0 = \frac{V_0}{6H_0^2}
\end{equation}
to solve Eq.(\ref{equ:eom2}) numerically. The evolution of $\psi$ and $w$ and their derivatives are plotted in Figs.\ref{fig:phiz3}-\ref{fig:wwz3}. 

\section{Dynamical analysis  }\label{sec:dyn}

Dynamical analysis is an effective method to reveal the novel phenomena arising from nonlinear equations without solving them. It can produce good numerical estimates of parameters connected with  general features such as stability. This method has already been used for analyzing the evolution of the universe, see Refs.\cite{Feng:2012wx}\cite{Feng:2014fsa}.   

In this section, we will treat the $f(\sigma)$ as a general function, which may have multiple zero points, and perform the dynamical analysis on the whole system of equations.

\subsection{Dyanamical equations}
From Eq.(\ref{equ:eom}) and using $d\phi = kd\sigma/f$, we have
\begin{equation}\label{equ:sig}
\ddot \sigma-\frac{df}{fd\sigma}\dot \sigma^2+3H\dot \sigma + \frac{dV}{d\sigma}\frac{f^2}{k^2} = 0\,.
\end{equation}
After defining the dimensionless variables in the units of $8\pi G= 1$:
\begin{eqnarray}\label{equ:def}
X=\frac{k\dot\sigma}{\sqrt{6}fH}\,,\quad Y = \frac{\sqrt{V}}{\sqrt{3} H} \,,\quad \lambda = \frac{f}{k}\frac{V_\sigma}{V}\,,
\end{eqnarray}
with potential $V_\sigma=dV/d\sigma$, we have the constraint arisen from the Friedmann equation
\begin{equation}
1 = \Omega_m + X^2+Y^2 \,, \quad \Omega_m = \frac{\rho_m}{3H^2}\,,
\end{equation}
therefore, the whole dynamical system is given by
\begin{eqnarray}
\frac{dX}{dx}&=& -3X- \sqrt{\frac{3}{2}} \lambda Y^2+  \frac{3}{2}X(1+X^2-Y^2) \,,\label{equ:sys1}\\
\frac{dY}{dx}&=&Y\left[ \sqrt{\frac{3}{2}} \lambda X +\frac{3}{2}(1+X^2-Y^2)\right]\,, \label{equ:sys2}\\
\frac{d\lambda}{dx} &=&   \sqrt{6}X\lambda \left( \Gamma -\lambda \right)\,,\label{equ:sys3}
\end{eqnarray}
where
\begin{equation}
\Gamma \equiv \frac{f_\sigma}{k} + \lambda \frac{VV_{\sigma\sigma}}{ V_\sigma^2}\,,
\end{equation}
with the convention $f_\sigma=df/d\sigma, V_{\sigma\sigma}=d^2V/d\sigma^2$. 
The equation of state can also be written in terms of $X,Y$:
\begin{equation}
w = \frac{X^2-Y^2}{X^2+Y^2}\,.
\end{equation}

Generally, the system (\ref{equ:sys1})-(\ref{equ:sys3}) is not an strictly autonomous system, but in some cases it is indeed an autonomous system. For example, when $\Gamma = \lambda$. In this case, it implies
\begin{equation}
\frac{df/d\sigma}{f} = \frac{dV/d\sigma}{V}-\frac{d^2V/d\sigma^2}{dV/d\sigma}\,.
\end{equation}
After integrating the above equation, we have $f\frac{dV}{d\sigma}\sim V$ up to some integration constant. By using the transformation
\begin{equation}
\phi = \int \frac{d\sigma}{f} \sim \int \frac{dV}{Vd\sigma}d\sigma = \ln V\,,
\end{equation}
we then get an exponential potential for $\phi$. Take $V=m^2\sigma^2/2$, we get $f\sim \sigma$, which corresponds to the $\beta=0$ case in Eq.(\ref{equ:mod1}). When $\Gamma=\lambda$, it shows $\lambda=\lambda_c$ is a constant due to Eq.(\ref{equ:sys3}). The dynamical system then reduces to $2-$dimensional one. 
When $\lambda_c\neq 0$,  there are five critical points ($X_c,Y_c$) 
\begin{eqnarray}\label{equ:5cr}
(0,0)\,, (1,0)\,,(-1,0) \,,\left(-\frac{\lambda_c}{\sqrt{6}},\sqrt{1-\frac{\lambda_c^2}{6}}\right)\,,\left(-\sqrt{\frac{3}{2}}\frac{1}{\lambda_c},\sqrt{\frac{3}{2}}\frac{1}{\lambda_c}\right)\,.
\end{eqnarray}
These critical points are the same as those for a quintessence model, and their stabilities have  already been investigated in the literature, see Ref.\cite{Bahamonde:2017ize} and references therein. 

When $\Gamma\neq \lambda$, Eqs.(\ref{equ:sys1})-(\ref{equ:sys3}) do not formulate an autonomous system. By constructing, the $f$ function has multi-zero points. When  the system approaches one of the zero points, $\lambda$  will become nearly vanishing due to the definition of  $\lambda\sim f$ in Eq.(\ref{equ:def}). The derivative of $\Gamma$ with respect to $x$ is given by 
\begin{equation}
\frac{d\Gamma}{dx} =\frac{f_{\sigma\sigma}f}{k^2}\sqrt{6}X + \frac{d\lambda}{dx} \frac{VV_{\sigma\sigma}}{ V_\sigma^2} + \lambda \frac{d}{dx}\left( \frac{VV_{\sigma\sigma}}{ V_\sigma^2}\right)\,.
\end{equation}
When $f\rightarrow0$, we have $\lambda\sim 0$, $d\lambda/dx\sim 0$ due to Eq.(\ref{equ:sys3}) and correspondingly $d\Gamma/dx\sim 0$ due to the above equation.

By introducing the following variables: 
\begin{eqnarray}
\Gamma_{A(1)} &=& \frac{f_\sigma}{k}\\
\Gamma_{A(n)} &=& \frac{f^{(n)} f^{n-1}}{f_\sigma^{n}} \,, \quad 
\Gamma_{B(n)} = \frac{V^{(n)} V^{n-1}}{V_\sigma^{n}} \,, n\geq 2\,,
\end{eqnarray}
where $f^{(n)} \equiv d^nf/d\sigma^n$ and $V^{(n)} \equiv d^nV/d\sigma^n$. We can rewrite $\Gamma$ into two parts:
\begin{equation}
\Gamma = \Gamma_{A(1)} + \lambda\Gamma_{B(2)} \,;
\end{equation}
therefore, the dynamical equations for these variables are given by 
\begin{eqnarray}
\frac{d\Gamma_{A(1)}}{dx} &=& \frac{f^{(2)}}{k}\frac{\dot\sigma}{H} = \sqrt{6}X\frac{f^{(2)}f}{k^2} = \sqrt{6}X\Gamma_{A(1)}^2\frac{f^{(2)}f}{f_\sigma^2} = \sqrt{6}X\Gamma_{A(1)}^2\Gamma_{A(2)} \,,\label{equ:sys4}\\
\frac{d\Gamma_{A(2)}}{dx} &=& \left( \frac{f^{(3)}f}{f_\sigma^2} + \frac{f^{(2)}f_\sigma}{f_\sigma^2}-2\frac{(f^{(2)})^2f}{f_\sigma^3}\right) \frac{\dot\sigma}{H} =\sqrt{6}X\Gamma_{A(1)} \bigg(\Gamma_{A(3)}+\Gamma_{A(2)}-2\Gamma_{A(2)}^2\bigg)\,,
\end{eqnarray}
and 
\begin{eqnarray}
\nonumber
\frac{d\Gamma_{A(n)}}{dx} &=& \left(\frac{f^{(n+1)} f^{n-1}}{f_\sigma^{n}}+(n-1)\frac{f^{(n)} f^{n-2}}{f_\sigma^{n-1}}-n\frac{f^{(n)} f^{n-1} f^{(2)}}{f_\sigma^{(n+1)}}\right)\frac{\dot\sigma}{H} \\
&=&\sqrt{6}X\Gamma_{A(1)}\bigg(\Gamma_{A(n+1)}+(n-1)\Gamma_{A(n)}-n\Gamma_{A(n)}\Gamma_{A(2)}\bigg)\,,\label{equ:sys5}\\
\nonumber
\frac{d\Gamma_{B(n)}}{dx} &=&  \left( \frac{V^{(n+1)} V^{n-1}}{V_\sigma^{n}} +  (n-1)\frac{V^{(n)} V^{n-2}}{V_\sigma^{(n-1)}} - n \frac{V^{(n)} V^{n-1}V^{(2)}}{V_\sigma^{(n+1)}} \right) \frac{\dot\sigma}{H} \\
&=& \sqrt{6}X\lambda\bigg(\Gamma_{B(n+1)}+  (n-1)\Gamma_{B(n)} - n \Gamma_{B(n)}\Gamma_{B(2)}\bigg)\,,\label{equ:sys6}
\end{eqnarray}
for $n=2,3,\cdots, N$. 

Note that  $f=0$ leads to $\lambda=0$, we then get $\Gamma_{A(n)}=0$,  and  $d\Gamma_{A(n)}/dx=d\Gamma_{B(n)}/dx=0$, $n=2,3,\cdots,N$, which is indeed a critical point for the whole $2(N+1)$-dimensional dynamical system Eqs.(\ref{equ:sys1})-(\ref{equ:sys3}), Eq.(\ref{equ:sys4}) and Eqs.(\ref{equ:sys5})(\ref{equ:sys6}).  The critical points projected on the subspace ($X_c,Y_c, \lambda_c=0 $) are
\begin{eqnarray}
(0,0,0)\,, (0,1,0)\,, (1,0,0)\,, (-1,0,0)\,, 
\end{eqnarray}
with constants $\Gamma_{A(1)c}, \Gamma_{A(n)c}$, and $\Gamma_{B(n)c}$. When $\lambda\neq 0$, there are three critical points,
\begin{eqnarray}
(0,0,\lambda_c)\,,\left(-\frac{\lambda_c}{\sqrt{6}},\sqrt{1-\frac{\lambda_c^2}{6}},\lambda_c\right)\,,\left(-\sqrt{\frac{3}{2}}\frac{1}{\lambda_c},\sqrt{\frac{3}{2}}\frac{1}{\lambda_c},\lambda_c\right)\,,
\end{eqnarray}
where the second point requires $\lambda_c^2\leq 6$. Both of the last two points require
\begin{eqnarray}
\Gamma_{A(1)} &=& 0\,,\\
\Gamma_{B(2)} &=& 1\,,\\
\Gamma_{B(n+1)}&=&\Gamma_{B(n)}\bigg[ n \Gamma_{B(2)}-  (n-1)\bigg] = \Gamma_{B(n)} = 1\,,\quad n\geq 2\,.\label{equ:bb}
\end{eqnarray}
In other words, these two points require an exponential potential that we have discussed before.

\subsection{Perturbations around the critical points}
When the critical points have $\lambda_c=0$, the linear perturbations of $\Gamma_{A(n)}$ are governed by
\begin{eqnarray}
\frac{d\delta\Gamma_{A(1)}}{dx} &=&\sqrt{6}X\Gamma_{A(1)}^2\delta\Gamma_{A(2)}\,,\\
\frac{d\delta \Gamma_{A(n)}}{dx} &=& \sqrt{6}X\Gamma_{A(1)}\bigg(\delta \Gamma_{A(n+1)}+(n-1)\delta \Gamma_{A(n)}\bigg)\,,\quad n\geq 2 \,,
\end{eqnarray}
and those of $\Gamma_{B(n)}$ are 
\begin{equation}
\frac{d\delta \Gamma_{B(n)}}{dx}
= 0\,,\quad n\geq 2\,.
\end{equation}
We also have 
\begin{equation}
\frac{d\delta \lambda}{dx}= \sqrt{6}X  \Gamma_{A(1)} \delta \lambda\,,\label{equ:per3}
\end{equation}
and
\begin{eqnarray}
\frac{d\delta X}{dx}&=& -3\delta X- \sqrt{\frac{3}{2}} Y^2\delta \lambda  +  \frac{3}{2}\delta X(1+X^2-Y^2) + 3(X\delta X-Y\delta Y) \,,\label{equ:sys11}\\
\frac{d\delta Y}{dx}&=& \frac{3}{2}(1+X^2-Y^2)\delta Y + Y\left[ \sqrt{\frac{3}{2}} \delta\lambda X  +3(X\delta X-Y \delta Y)\right]\,, \label{equ:sys12}
\end{eqnarray}
These perturbations $\delta \Gamma_{A(n)}$, $\delta \Gamma_{B(n)}$ and $\delta\lambda$ are obviously constants near the critical points ($0,0,0$) and ($0,1,0$). Eqs.(\ref{equ:sys11})(\ref{equ:sys12}) become
\begin{eqnarray}
\frac{d\delta X}{dx}&=& - \frac{3}{2}\delta X  \,,\\
\frac{d\delta Y}{dx}&=& \frac{3}{2}\delta Y 
\end{eqnarray}
near the critical point $(0,0,0)$ and 
\begin{eqnarray}
\frac{d\delta X}{dx}&=& -3\delta X- \sqrt{\frac{3}{2}} \delta \lambda  -3\delta Y \,,\\
\frac{d\delta Y}{dx}&=&  - 3\delta Y
\end{eqnarray}
near the critical point $(0,1,0)$. 

The critical point $(0,0,0)$ corresponds to the matter dominated universe with $\Omega_m=1$, and it is a saddle point; while the critical point $(0,1,0)$ corresponds to the de Sitter universe, in which the potential of $\phi$ dominates the energy density. From Eq.(\ref{equ:per3}), $d\delta\lambda/dx = 0 $, so it leads to a vanished determinant of the coefficient matrix for the linear perturbation system.

\begin{figure}[h]
	\begin{center}
		\includegraphics[width=0.4\textwidth,angle=0]{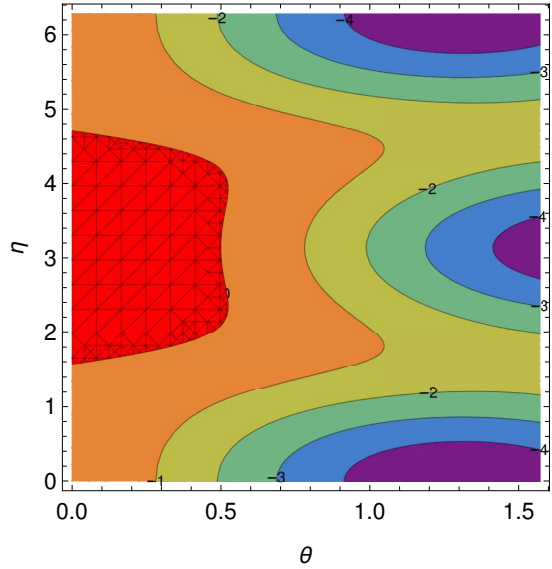}
		\caption{\label{fig:rfun} The space of ($\theta,\eta$). The red gridding range corresponds to $R(\theta,\eta)>0$, while others correspond to $R(\theta,\eta)<0$. }
	\end{center}
\end{figure}

Let $X = r\sin\theta\cos\eta, \sqrt{1-Y^2} = r\sin\theta\sin\eta, \lambda = r\cos\theta$, then we have 
$X^2+1-Y^2+\lambda^2 = r^2$. The critical point ($0,1,0$) corresponds to $r=0$. 
The dynamical system (\ref{equ:sys1})-(\ref{equ:sys3}) become
\begin{eqnarray}
\frac{dr}{dx} &=&r R(\theta,\eta) + o(r)\,,\label{equ:d1}\\
\frac{d\theta}{dx} &=& R(\theta,\eta) \cot\theta + o(r)\,,\label{equ:d2}\\
\frac{d\eta}{dx} &=& \Xi(\theta,\eta) + o(r)\label{equ:d3}
\end{eqnarray}
with
\begin{eqnarray}
R(\theta,\eta) &\equiv& -\frac{1}{2}\bigg[ \left(\sqrt{6}\sin\theta\cos\eta+\cos\theta\right)^2  +4\sin^2\theta - 1 \bigg] \,,\\
\Xi(\theta,\eta) &\equiv& -\frac{1}{2}\cos (2\eta) \csc \eta \left(\sqrt{6} \cot \theta+3 \cos \eta\right)\,.
\end{eqnarray}
Then we have
\begin{eqnarray}
\frac{dr}{rd\theta} &=& \tan\theta \,, \label{equ:dr} \\
\frac{dr}{rd\eta} &=& \frac{R(\theta,\eta)}{\Xi(\theta,\eta)} 
= \frac{\sin \eta \left(\sin ^2\theta (6 \sec (2 \eta)+3)+\sqrt{6} \sin (2 \theta ) \cos  \eta \sec (2 \eta)\right)}{\sqrt{6} \cot \theta +3 \cos\eta}\,.\label{equ:dr2}
\end{eqnarray}
Eq.(\ref{equ:dr}) indicates $r$  increases when $\theta$ become large; therefore, if $\theta$ decreases with time,that is $d\theta/dx<0$, $r$ will decrease, the system is thus stable as an attractor at the point $(0,1,0)$. Actually, from Eq.(\ref{equ:d1}), $r$ will decrease with time when $R(\theta,\eta)<0$. That the range of ($\theta,\eta$)  makes $R(\theta,\eta)<0$ is plotted in Fig.\ref{fig:rfun}, in which it is the part without gridding. 

The critical point ($\pm 1, 0, 0$) corresponds to the universe dominated by the kinetic energy of $\phi$, and the perturbations of $X,Y,\lambda $  around this point are governed by 
\begin{eqnarray}
\frac{d\delta X}{dx}&=&  3X\delta X \,,\\
\frac{d\delta Y}{dx}&=& 3\delta Y \,,\\
\frac{d\delta \lambda}{dx} &=&   \sqrt{6} X  \Gamma_{A(1)} \delta \lambda  \,.
\end{eqnarray}
These two points are both unstable critical points. In the case of $\lambda_c\neq 0$, the critical point ($0, 0, \lambda_c $) is not interesting, for it corresponds to the universe without $\Omega_\phi$, and it is a saddle point just like  the ($0, 0, 0$) point. The other two points have  already been investigated in the literature.  In summary, the multi-pole dark energy model does have stable attractor solutions just as the quintessence model.

\section{Discussions and Conclusions}\label{sec:dis}

In the multi-pole dark energy model, a flat potential for the field $\sigma$ is no longer needed. After transforming to the canonical kinetic form, we could have a stable solution, which corresponds to the dark energy dominated universe. A scaling solution could be also obtained. For example, if $V(\sigma)=m^2\sigma^2/2$, and the required potential of $\psi$ that leads to a constant equation of state $w=w_c$ is $V_s(\psi)$, then function $f$ should be chosen as 
\begin{equation}
f(\sigma) = \sqrt{\frac{1}{2m^2V_s}} \frac{dV_s}{d\phi}\bigg(\phi = V_s^{-1}(m^2\sigma^2/2) \bigg)\,,
\end{equation}  
where $V_s^{-1}$ is the inverse function of $V_s$. 

The whole  dynamical system Eqs.(\ref{equ:sys1})-(\ref{equ:sys3}), Eq.(\ref{equ:sys4}) and Eqs.(\ref{equ:sys5})(\ref{equ:sys6}) seems to have infinite dimensions, since there is always a new variable $\Gamma_{A(n+1)}$ or $\Gamma_{B(n+1)}$ that appears in the equation of $d\Gamma_{A(n)}/dx$ or $d\Gamma_{B(n)}/dx$. If the function $f$ or $V$ has a maximum order of $\sigma$, e.g. $\sigma^{N-1}$, then $\Gamma_{A(N)}=0$ or $\Gamma_{B(N)}=0$. As a result, the whole system is closed to form an autonomous system, and it has $2(N+1)$-dimensions.

In conclusion, we have proposed a multi-pole dark energy model. The cosmological evolution is obtained explicitly for the two pole model, while dynamical analysis on the whole system is performed for the multi-pole model. We find that this kind of dark energy model could have a stable solution, which corresponds to the universe dominated by the potential energy of the scalar field. Thus, the multi-pole dark energy also appears worthy of future investigation.

\acknowledgments
This work is supported by National Science Foundation of China grant Nos.~11105091 and~11047138, ``Chen Guang" project supported by Shanghai Municipal Education Commission and Shanghai Education Development Foundation Grant No. 12CG51, and Shanghai Natural Science Foundation, China grant No.~10ZR1422000. CJF would like to thank Prof. Eric V. Linder for very helpful comments.

\end{document}